\documentstyle{mn}
\input{epsf}

\def\gtt{g_{tt}}
\def\gtp{g_{t\phi}}
\def\gpp{g_{\phi \phi}}
\def\ct{\cos \theta_e}

\def\pem{\phi_e}

\title[Relativistic accretion disc]
{Application of a relativistic accretion disc model to X-ray spectra of LMC X-1 and 
GRO J1655--40}

\author[M. Gierli\'nski, A. Macio{\l}ek-Nied{\'z}wiecki and K. Ebisawa]
{Marek~Gierli\'nski$^{1,2}$, Andrzej~Macio{\l}ek-Nied{\'z}wiecki$^{3}$ and 
Ken~Ebisawa$^{4}$\\
$^1$Astronomical Observatory, Jagiellonian University, Orla 171, 30-244 Cracow,
Poland\\
$^2$N. Copernicus Astronomical Center, Bartycka 18, 00-716 Warsaw, Poland\\
$^3${\L}{\'o}d{\'z} University, Department of Physics, Pomorska 149/153, 90-236 
{\L}{\'o}d{\'z}, Poland\\
$^4$Code 660.2, Laboratory for High Energy Astrophysics, NASA/Goddard Space
Flight Center, Greenbelt, MD 20771, USA\\(also at Universities Space Research
Association)\\}

\date{Accepted on 2001 March 21}
\pagerange{\pageref{firstpage}--\pageref{lastpage}}
\pubyear{2001}

\begin{document}

\topmargin = -0.5cm

\maketitle

\label{firstpage}

\begin{abstract}
  
  We present a general relativistic accretion disc model and its
  application to the soft-state X-ray spectra of black hole binaries.
  The model assumes a flat, optically thick disc around a rotating
  Kerr black hole. The disc locally radiates away the dissipated
  energy as a blackbody. Special and general relativistic effects
  influencing photons emitted by the disc are taken into account. The
  emerging spectrum, as seen by a distant observer, is parametrized by
  the black hole mass and spin, the accretion rate, the disc
  inclination angle and the inner disc radius.
  
  We fit the {\it ASCA\/} soft state X-ray spectra of LMC X-1 and GRO
  J1655--40 by this model. We find that having additional limits on
  the black hole mass and inclination angle from optical/UV
  observations, we can constrain the black hole spin from X-ray data.
  In LMC X-1 the constrain is weak, we can only rule out the maximally
  rotating black hole. In GRO J1655--40 we can limit the spin much
  better, and we find $0.68 \leq a \leq 0.88$. Accretion discs in both
  sources are radiation pressure dominated.  We don't find Compton
  reflection features in the spectra of any of these objects.

\end{abstract}


\section{Introduction}
\label{sec:introduction}

Black hole binaries (BHB) are generally observed in one of the two spectral states, 
as determined by their X-ray/$\gamma$-ray energy spectra. In the hard state most of 
energy is radiated above $\sim$ 10 keV. The spectrum is dominated by a power law with 
the photon spectral index $\Gamma < 2.0$ and a high-energy cutoff above 100 keV 
(Grove et al.\ 1998). A common interpretation of this emission is in terms of 
Comptonization of soft seed photons by thermal, hot, optically-thin electron plasma. 
Cyg X-1 (Gierli{\'n}ski et al.\ 1997) and GX339--4 (Zdziarski et al.\ 1998) are 
typical examples of the hard state BHB. In the soft state most of the energy is 
radiated in the soft X-rays, below $\sim$ 10 keV. This energy range is dominated by a 
blackbody-like component with characteristic temperature $\sim$ 1 keV. This component 
is usually attributed to the thermal emission of a cold, optically-thick accretion 
disc, extending down to the marginally stable orbit (Shakura \& Sunyaev 1973). The 
disc spectrum is often accompanied by a high-energy tail, which can be described as a 
power law with a typical photon spectral index $\Gamma \sim 2.0$--$2.5$, extending 
into $\gamma$-rays without apparent break or cutoff (Grove et al.\ 1998). LMC X-1 
(Schlegel et al.\ 1994), LMC X-3 (Ebisawa et al.\ 1993), GS 1124--68 just after the 
outburst (Ebisawa et al.\ 1994) or Cyg X-1 between May and August 1996 
(Gierli{\'n}ski et al.\ 1999) are the examples of the soft state BHB.

At first approximation the soft component in the soft state can be
described by a multicolour disc model (hereafter MCD; Mitsuda et al.\ 
1984), which spectrum is a sum of blackbodies with radial temperature
distribution $T(r) \propto r^{-3/4}$. This model has been commonly
used for spectral fitting by many authors. Though MCD model
approximates the disc spectral shape well, it ignores the inner
torque-free boundary condition and parameters derived through it are
incorrect. An improvement of the MCD model is a multicolour disc model
in the pseudo-Newtonian potential (Gierli{\'n}ski et al.\ 1999),
taking properly into account the boundary condition.

In vicinity of the black hole relativistic effects become important. The effects of 
relativity on the emerging disc spectrum have been studied by Cunningham (1975), 
Laor, Netzer \& Piran (1990), Asaoka (1989), Hanawa (1989), Fu \& Taam (1990), Yamada 
et al.\ (1994)  and others. Hanawa (1989) calculated disc spectra around a 
non-rotating black hole, creating a fast-computing model that have been used for 
spectral fitting several times (e.g.\ Ebisawa, Mitsuda \& Hanawa 1991; Makishima et 
al.\ 2000). However, this model is limited to Schwarzschild metric only and effects 
of light bending have been neglected. More advanced model was developed by Laor et 
al.\ (1990), who studied accretion discs around rotating black hole, taking into 
account both relativistic effects affecting radiation and vertical structure of the 
disc. Laor (1990) applied this model to the AGN spectra.

In this paper we show an application of the general relativistic (hereafter GR) disc 
model to the soft state X-ray spectra of BHB. We focus on estimating the black hole 
spin. The model thoroughly treats relativistic effects in the deep gravitational 
potential of the black hole. In Section \ref{sec:model} we give description of the 
model and show examples of the model spectra. Then, we fit the X-ray spectral data of 
LMC X-1 (Section \ref{sec:lmc_x-1}) and GRO J1655--40 (Section 
\ref{sec:gro_j1655-40}) with the GR model. In Section \ref{sec:discussion} we discuss 
the obtained results.

\section{Model description}
\label{sec:model}

We consider an accretion disc around a black hole (e.g.\ Page \&
Thorne 1974). The model is based upon certain assumptions. We assume
that the disc rotates in the equatorial plane around the Kerr black
hole; that it is in a steady state; that it is axially symmetric; that
it is flat and geometrically thin; that it is optically thick; that it
locally radiates away the dissipated gravitational energy as a
blackbody; that there is no emission below the marginally stable
orbit.

A black hole is characterized by its mass, $M$, and angular momentum, $J$. We use 
dimensionless spin $a \equiv Jc/GM^2$, where $0 \leq a \leq 1$ and $a = 0$ 
corresponds to the non-rotating (Schwarzschild) black hole. In accreting systems the 
radiation emitted by the disc produces a counteracting torque and the black hole 
cannot be spun-up beyond $a = 0.998$ (Thorne 1974). Therefore we do not exceed this 
value in our model and refer to it as to the maximally rotating black hole. The 
efficiency of accretion, $\eta$, varies from $\eta \approx 0.057$ for the 
non-rotating black hole to $\eta \approx 0.32$ for $a = 0.998$.

The spectrum of the disc, $F_\nu$, seen at infinity is computed by means of a photon 
transfer function, $\cal T$, which describes travel of photons from the point of 
origin to the distant observer. The photon of frequency $\nu_e$ is emitted at radius 
$r_e$, at cosine angle $\mu_e \equiv \cos \theta_e$ and then perceived at frequency 
$\nu_o$ by the observer at cosine angle $\mu_o \equiv \cos i$. The observed flux is
\begin{eqnarray}
\lefteqn{F_{\nu_o}(\mu_o) = \left( R_g \over D \right)^2 \nu_o \! \int \!\! {\rm 
d}g_{\rm eff} \!\! \int \!\! {\rm d}r_e \!\! \int \!\! {\rm d}\mu_e \,} \nonumber \\
\lefteqn{~~~~\times r \,{\cal T}(a, r_e, \mu_e, g_{\rm eff}, \mu_o) \, N(r_e, \mu_e, 
{\nu_o \over g_{\rm eff}}).}
\label{eq:trans_integ}
\end{eqnarray}
Subscript `$e$' denotes quantities measured in the local frame co-moving with the 
disc and subscript `$o$' denotes quantities measured by the observer at infinity. 
$N(r_e, \mu_e, {\nu_o \over g_{\rm eff}}) \equiv N_{\nu_e}(r_e, \mu_e)$ is the 
specific photon number intensity of the disc emission, $g_{\rm eff} \equiv \nu_o / 
\nu_e$ is the effective redshift of the photon, $D$ is the distance to the source and 
$R_g \equiv GM/c^2$ is the gravitational radius. Numerical constants are included in 
the transfer function.

The transfer function treats the special and general relativistic effects affecting 
the spectrum. It takes into account the Doppler energy shift from the fast moving 
matter in the disc, the gravitational shift and the light bending near the massive 
object. It includes integration over the azimuthal angle $\phi_e$. Each element of 
the transfer function, ${\cal T}(a, r_e, \mu_e, g_{\rm eff}, \mu_o)$, was computed by 
summing all photon trajectories for all angles $\phi_e$, at given ($r_e$, $\mu_e$), 
for which required ($g_{\rm eff}$, $\mu_o$) were obtained. The details of the 
transfer function computation are given in Appendix \ref{app:transfer_function}. A 
similar approach to computing the black hole disc spectra has been applied, e.g., by 
Laor et al.\ (1990) and Asaoka (1989). Note that our construction of  $\cal T$ is 
slightly different than that of Cunningham (1975), as the latter one involves the 
relation of the specific intensities at the emitter and the observer, respectively, 
given by Liouville's theorem, while our model is based on counting individual 
photons. 

The local gravitational energy release per unit area of the disc, per unit time is 
(Page \& Thorne 1974)
\begin{eqnarray}
\lefteqn{Q(x)  = {3 \dot{M}_{\rm d} c^6 \over 8 \pi G^2 M^2} \, {1 \over x^4(x^3 - 3x 
+ 2a)} \left[ x - x_0 - {3 \over 2} a \ln \left({x \over x_0}\right) \right.} 
\nonumber \\
& & - {3(x_1 - a)^2 \over x_1(x_1 - x_2)(x_1 - x_3)} \ln \left({x - x_1 \over x_0 - 
x_1}\right) \nonumber \\
& & - {3(x_2 - a)^2 \over x_2(x_2 - x_1)(x_2 - x_3)} \ln \left({x - x_2 \over x_0 - 
x_2}\right) \nonumber \\
& & \left. - {3(x_3 - a)^2 \over x_3(x_3 - x_1)(x_3 - x_2)} \ln \left({x - x_3 \over 
x_0 - x_3}\right) \right],
\end{eqnarray}
where $x = r_e^{1/2}$, $x_0 = r_{\rm ms}^{1/2}$, $x_1 = 2 \cos({1 \over 3} \arccos a 
- \pi / 3)$, $x_2 = 2 \cos({1 \over 3} \arccos a + \pi / 3)$ and $x_3 = -2 \cos({1 
\over 3} \arccos a)$. 
\begin{equation}
r_{\rm ms} = 3 + A_2 - {\rm sign} \, a \, \left[(3 - A_1)(3 + A_1 + 
2A_2)\right]^{1/2}
\label{eq:rms}
\end{equation}
is the marginally stable orbit radius, where $A_1 = 1 + (1 - a^2)^{1/3}[(1 + a)^{1/3} 
+ (1 - a)^{1/3}]$ and $A_2 = (3a^2 + A_1^2)^{1/2}$. Both $r_e$ and $r_{\rm ms}$ are 
expressed in units of $R_g$.

The photon number intensity, $N_{\nu_e}$, can be derived from the energy release 
rate, $Q$. However, the relation between these two quantities is not simply 
$Q(r_e)=\int h \nu_e N_{\nu_e}(r_e,\mu_e) {\rm d} \nu_e {\rm d} \mu_e$, as assumed in 
previous similar calculations (e.g., Laor et al.\ 1990). The photon number intensity 
which is used in equation (\ref{eq:trans_integ}), is defined in terms of the distant 
observer coordinate frame. On the other hand, $Q$ is defined as the energy 
dissipated per unit proper time per unit proper area, as measured by the observer 
orbiting with the disc. The corresponding dissipation rate measured by the distant 
observer will be affected by kinematic and gravitational time dilation and length 
contraction. We emphasize that these effects are not included in the calculation of 
the single photon trajectory, and are treated separately when applying the transfer 
function formalism.

In Appendix \ref{app:intensity_transformation} we derive transformation of the time 
and the disc surface area between the disc rest frame and the frame of the distant 
observer:
\begin{equation}
{\rm d}t'=\beta_t {\rm d}t,~~~~~~~~~~~{\rm d} r' {\rm d} \phi'=\beta_S {\rm d} r {\rm 
d} \phi,
\end{equation}
where $(t,r,\phi)$ and $(t',r',\phi')$ are the coordinates of the reference frame of 
the distant observer and the disc, respectively.

In order to find a locally emitted spectrum we would need to compute the vertical 
structure of the disc. However, this problem did not find a satisfactory general 
solution yet. Instead, we simply assume that each point of the disc radiates like a 
blackbody and introduce two corrections. First, we note that the Thomson scattering 
dominates over absorption in the inner part of the disc and the local spectrum is 
affected by Comptonization. At a given radius, $r_e$, we approximate the spectrum by 
a diluted blackbody,
\begin{eqnarray}
B_{\nu_e}^{\rm db} = {1 \over f_{\rm col}^4} B_{\nu_e}(f_{\rm col} T_{\rm eff}),
\end{eqnarray}
where $B_{\nu_e}$ is the Planck function, $f_{\rm col} = T_{\rm col}/T_{\rm eff}$ is the 
spectral hardening factor, $T_{\rm col}$ is locally observed colour temperature and 
$T_{\rm eff}$ is the effective temperature, related to the locally dissipated energy, 
$Q$, as $\sigma T_{\rm eff}^4 = Q$. Following Shimura \& Takahara (1995) we use 
$f_{\rm col} = 1.7$. 

The second correction concerns angular distribution of local emission. We assume a 
linear limb darkening in form:
\begin{eqnarray}
I_{\nu_e}(\mu_e) = I_0 {1 + \delta\mu_e \over  1 + \delta} B_{\nu_e}^{\rm db},
\end{eqnarray}
where $\delta = 2.06$ in classical electron scattering limit (e.g.\ Phillip \& 
M{\'e}sz{\'a}ros 1986). $I_{\nu_e}$ is the specific intensity in the local frame, 
co-rotating with the disc. Factor $I_0$ is found from the requirement of energy 
conservation. Namely, the power emitted by a limb-darkened surface element d$A$, 
$2\pi \int_0^1 I_{\nu_e}(\mu_e) \mu_e $d$\mu_e$d$A$, should be equal to the power 
emitted by the same surface element without limb darkening, $\pi B_{\nu_e}^{\rm 
db}\,$d$A$. From this we find
\begin{eqnarray}
I_0 = 3 {1 + \delta \over 3 + 2\delta},
\end{eqnarray}
and
\begin{eqnarray}
I_{\nu_e}(\mu_e) = 3{1 + \delta\mu_e \over  3 + 2\delta} B_{\nu_e}^{\rm db}.
\end{eqnarray}

Then, the specific photon number intensity of the disc, measured by the distant 
observer, is given by
\begin{equation}
N_{\nu_e}(\mu_e) = \beta_S \beta_t {I_{\nu_e}(\mu_e) \over h \nu_e}.
\end{equation}

\begin{figure}
\begin{center}
\leavevmode
\epsfxsize=8cm \epsfbox{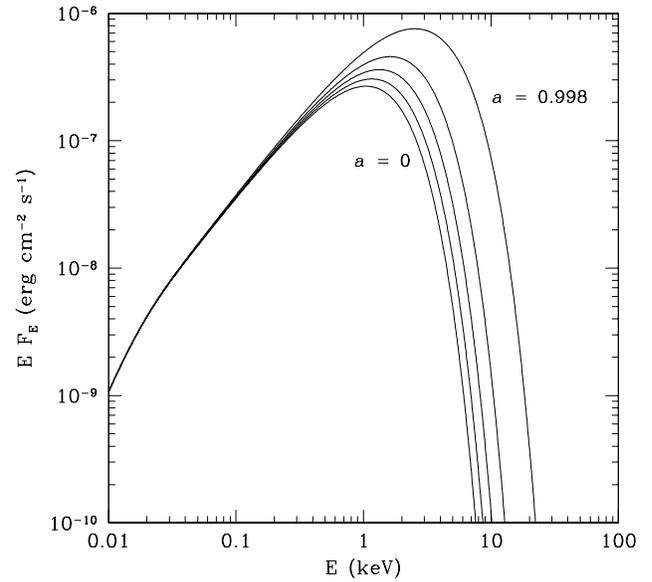}
\end{center}
\caption{Spectra from the GR disc model for $a$ = 0, 0.25, 0.5, 0.75 and 0.998 (from the 
bottom to the top). The disc extends from the marginally stable orbit to $r_{\rm out} 
= 10^4$. The spectra were computed for a 10M$_\odot$ black hole at 1 kpc, with 
accretion rate $\dot{M}_{\rm d} = 10^{18}$ g s$^{-1}$, disc inclination angle $i = 
45^\circ$, $f_{\rm col}$ = 1.7 and $\delta$ = 2.06.}
\label{fig:disc_a}
\end{figure}

\begin{figure}
\begin{center}
\leavevmode
\epsfxsize=8cm \epsfbox{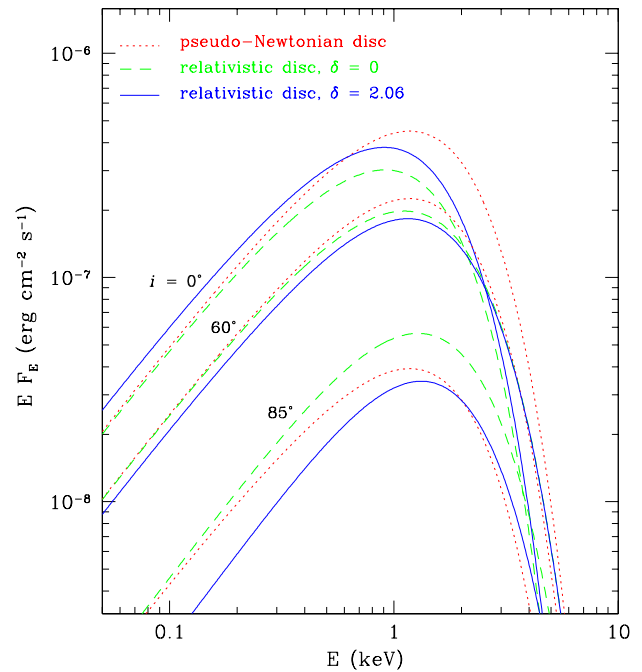}
\end{center}
\caption{Comparison of the GR and PN disc models for different inclination angles. 
For both discs $r_{\rm in} = 6$. The spectra were computed for a non-rotating 
10M$_\odot$ black hole at 1 kpc, with accretion rate $\dot{M}_{\rm d} = 10^{18}$ g 
s$^{-1}$, $f_{\rm col}$ = 1.7. The effect of the limb darkening in the relativistic 
disc is also shown.}
\label{fig:discr_vs_pn}
\end{figure}

There is, however, one effect that we have not taken into account. Due
to gravitational focusing some of the emitted photons return to the
disc, where they are reprocessed or scattered. Since the returning
photons alter the locally emitted spectrum recurrently, taking them
into account would require enormous computing time and would make
spectral fitting practically impossible. In our model returning
photons are lost, so the disc luminosity and temperature are
underestimated for the spin close to maximum. This effect was studied
by Cunningham (1976). He found that returning radiation is negligible
for $a < 0.9$. The difference between the actual locally generated
flux and that in the absence of returning radiation is only a few per
cent over most of the inner disc for the maximally rotating black
hole. At lower spins, the effect is much less pronounced. In this
paper we compute the GR disc spectra for $a$ = 0, 0.25, 0.5, 0.75 and
0.998. Though the $a = 0.998$ model underestimates temperature and
luminosity, the interpolation between the models would give accurate
results for $a < 0.9$, and slightly overestimated spin measurement
otherwise.

\begin{figure}
\begin{center}
\leavevmode
\epsfxsize=8cm \epsfbox{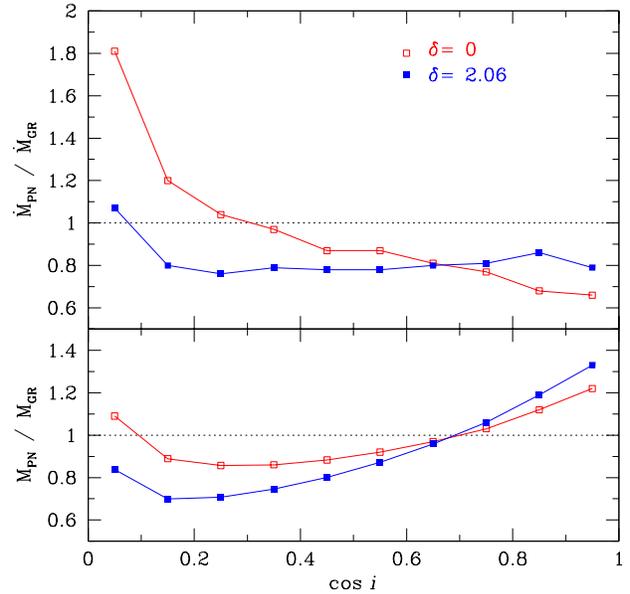}
\end{center}
\caption{Fits of the pseudo-Newtonian disc model to the fake spectra generated from the 
GR model. The data were created for a non-rotating 10M$_\odot$ 
black hole with $\dot{M}_{\rm d} = 10^{18}$ g s$^{-1}$ with (filled squares) and 
without (empty squares) limb darkening.}
\label{fig:pn_corr}
\end{figure}

Figure \ref{fig:disc_a} shows the GR disc spectra for different values
of the black hole spin. We clearly see how the accretion efficiency
grows with increasing spin, when energy is dissipated deeper in the
gravitational potential. On Figure \ref{fig:discr_vs_pn} we compare
spectra of the GR model (with $a = 0$) and the pseudo-Newtonian (PN)
model (Gierli{\'n}ski et al.\ 1999). The PN model approximates
temperature distribution with good accuracy, but effects of Doppler
and gravitational shifts, light bending and focusing are neglected and
light propagates through flat space. With increasing inclination angle
both energy of the peak and normalization of the relativistic disc
spectrum (dashed curve) increase in compare to the PN disc spectrum
(dotted curve), mostly due to Doppler shifts and increase of the
apparent disc area. A similar effect was found by Fu \& Taam (1990).
On the same figure we also show the effect of the limb darkening
(solid curve), enhancing the spectrum for lower and diminishing it for
higher inclination angles. Next, we check how the parameters obtained
with the PN model relate to these from the GR model. To do this, we
created fake {\it ASCA\/} spectra using the GR model with the
following parameters: $M = 10$M$_\odot$, $a = 0$, $\dot{M}_{\rm d} =
10^{18}$ g s$^{-1}$, $D$ = 1 kpc, $r_{\rm in} = 6$, $r_{\rm out} =
10^4$, $f_{\rm col} = 1.7$. The spectra were computed for $\delta = 0$
and 2.06, for several inclination angles. Then, we fitted the PN disc
model to the spectra finding the black hole mass and the accretion
rate. The results are presented on Figure \ref{fig:pn_corr}. If the
limb darkening is taken into account in the generated data, the
accretion rate obtained with the PN model is underestimated by factor
$\sim$ 0.8, almost independently of the inclination angle.  The mass
estimation is exact at inclination $i \approx 45^\circ$,
underestimated for higher and overestimated for lower inclination
angles. We stress that these fits were performed for a non-rotating
black hole only.

Zhang, Cui and Chen (1997a) investigated accretion onto a rotating
black hole in an approximate way. They used the MCD model and applied
GR corrections due to fractional change in the colour temperature and
due to change in the integrated flux. As a result, they were able to
constrain the black hole spin of a few BHB, including GRO J1655--40,
which we will analyze in details in Section \ref{sec:gro_j1655-40}.
For comparison, we have applied their method to the spectra created by
our GR model. The result depends on the GR corrections applied.  Since
Zhang et al.\ (1997) tabulated these corrections for $a$ = 0 and 0.998
only, the discrepancies are largest for $a \sim 0.5$. In particular,
for spectra generated by the GR model with $a$ = 0.998, 0.75 and 0.5,
the approximate treatment yields $a$ = 0.95, 0.58 and 0.24,
respectively. We however note, that with GR corrections calculated
precisely for all spin and inclination angle values the approximate
method can provide with reasonably accurate results.

The relativistic disc model parameters are: the black hole mass, $M$, and its spin, 
$a$, the inner and outer disc radii, $r_{\rm in}$ and $r_{\rm out}$, respectively (in 
units of $R_g$), the accretion rate, $\dot{M}_{\rm d}$ and the inclination angle, 
$i$. Subscript `d' denotes the accretion rate in the disc only, excluding all 
possible energy dissipation outside the disc. In our fits we always keep $r_{\rm out} 
= 10^4$, so the model has at most 5 free parameters. 

We underline that the overall model spectral shape does not change significantly, 
when these parameters vary. At first approximation, there are only two quantities 
fixing the spectrum: the energy of the peak in the spectrum (or the temperature) and 
the total flux (normalization). They translate into five model parameters, therefore 
only two out of them can be established uniquely. In most of our fits we fix $i$, $a$ 
and $r_{\rm in}$, leaving $M$ and $\dot{M}_{\rm d}$ as free parameters. As we show 
later, the inner disc radius as a free parameter can be also constrained to some 
extent.

However, the model, though unable to establish all parameters simultaneously, yields 
tight correlations between them. In particular, for a given spectrum and $r_{\rm in} 
= r_{\rm ms}$, as $a$ increases, $\dot{M}_{\rm d}$ decreases (to compensate the 
increased accretion efficiency) but $M$ increases (to keep the absolute inner disc 
radius $R_{\rm in} \propto r_{\rm in} M$ when $r_{\rm in}$ decreases). For a given 
spectrum and $a$ fixed, as $r_{\rm in}$ increases, $\dot{M}_{\rm d}$ increases but 
$M$ decreases. Having any additional restrictions for black hole mass and inclination 
angle we can hope to estimate the black hole spin.


\section{Application to the data}
\label{disc_data}

For spectral fits, we use {\sc xspec} v10 (Arnaud 1996). The confidence range of each 
model parameter is given for a 90 per cent confidence interval, i.e., $\Delta \chi^2= 
2.7$ (e.g.\ Press et al.\ 1992). 

The X-ray spectra of both sources can be decomposed into two
components: a soft, thermal emission peaking around a few keV and a
high-energy tail. We interpret them in terms of an optically thick
cold accretion disc and optically thin hot plasma.  Therefore we fit
the data by a model consisting of the GR disc and a high-energy tail,
which we model either by a power law or by thermal Comptonization (see
below).  The model spectra are absorbed by the interstellar medium
with a column density $N_{\rm H}$, for which we use the opacities of
Ba{\l}uci{\'n}ska-Church \& McCammon (1992) and the abundances of
Anders \& Grevesse (1989).

While constructing the disc model, we have tabulated the transfer function, $\cal T$, 
for five values of the black hole spin: $a$ = 0, 0.25, 0.5, 0.75 and 0.998. Since the 
radius of the marginally stable orbit, which is the lower limit of $r_{\rm in}$, is a 
function of $a$, we cannot interpolate between transfer functions tabulated for 
different values of $a$, when $r_{\rm in}$ is close to $r_{\rm ms}$. Therefore, we 
fit the data only for these five fixed values of $a$. In all the fits we keep the 
spectral hardening factor, $f_{\rm col} = 1.7$ and the limb darkening factor $\delta$ 
= 2.06. 


\subsection{LMC X-1}
\label{sec:lmc_x-1}

\subsubsection{Source properties}
\label{sec:lmc_x-1_properties}

LMC X-1 is a luminous X-ray source in the Large Magellanic Cloud. It had been a 
long-standing controversy about the optical counterpart of the X-ray source, since 
the position of LMC X-1 established by {\it Einstein} was between two stars separated 
by only 6". Cowley et al.\ (1995) improved the position of LMC X-1 from {\it 
ROSAT\/}-HRI observations and confirmed that the optical counterpart is a peculiar O7 
III Star \#32 with visual magnitude $V \sim 14.8$. Using optical spectroscopy 
Hutchings et al.\ (1987) showed that Star \#32 is in a binary system with orbital 
period of 4.2288 days. 

They found the mass function,  $f_M = 0.14\pm0.05$M$_\odot$, and the lower limit for 
the mass of the compact object, $M > 4$M$_\odot$, which makes LMC X-1 a good 
candidate for the stellar-mass black hole. The inclination angle of the binary is 
constrained by the lack of X-ray eclipses to be $i < 63^\circ.5$. The mass ratio, $q 
\equiv M_{\ast} / M$ is greater than 2, which, together with the mass function yields
\begin{equation}
M \sin^3 i > 0.8{\rm M}_\odot.
\end{equation}
The upper limit for the companion star is 25M$_\odot$, therefore $M < 12.5$M$_\odot$ 
and $i > 24^\circ$.

The distance to the Large Magellanic Cloud (LMC) is uncertain. It is usually 
published in terms of the distance modulus, $m - {\cal M} = 5\log D_{\rm pc} - 5$, 
where $m$ and ${\cal M}$ are apparent and absolute visual magnitudes and $D_{\rm pc}$ 
is the distance expressed in parsecs. Different determinations of the distance 
modulus result in conflicting results, from $18.065\pm0.12$ (Stanek, Zaritsky \& 
Harris 1998) to $18.7\pm0.1$ (Feast \& Catchpole 1997). This corresponds to the span 
in distance from 38.8 kpc to 57.5 kpc. Since the compact object mass determined from 
the disc observations depends linearly on $D$, this will significantly increase the 
uncertainty of mass determination.

The X-ray spectrum of LMC X-1 resembles the soft state spectrum of Cyg X-1 
(Gierli{\'n}ski et al.\ 1999). At first approximation it can be represented by a 
blackbody with temperature $\sim 0.7$ keV and a high-energy power-law tail beyond the 
blackbody (e.g.\ Ebisawa, Mitsuda \& Inoue 1989; Schlegel et al.\ 1994; Treves at 
al.\ 2000; Schmidtke, Ponder \& Cowley 1999; Nowak et al.\ 2001). No transition to 
the hard state has been ever observed.


\subsubsection{Spectral fits}
\label{sec:lmc_x-1_fits}

LMC X-1 was observed by {\it ASCA\/} on April 2--3, 1995. We have extracted the SIS 
spectrum with the net exposure time of 23.6 ks. In our fits we use the 0.7--10 keV 
spectrum, except the channels between 1.8--2.2 keV, which are strongly affected by 
the instrumental gold feature. Since we are interested in the shape of the disc 
continuum, this small gap in energy channels will not affect our results.

Both the distance to LMC X-1, $D$, and the inclination angle of the system, $i$, are 
uncertain. In most of our fits we use $D = 50$ kpc and $i = 50^\circ$, but we also 
check our results within the acceptable range of $38.8$ kpc $< D < 57.5$ kpc and 
$24^\circ < i < 63^\circ.5$. We do not impose any constraints on the black hole mass, 
$M$, during the fits, but check its consistency with the limits obtained from the 
optical observations, $4.0$M$_\odot \leq M \leq 12.5$M$_\odot$, {\it a posteriori}.

For the high-energy tail beyond the disc spectrum we chose a thermal Comptonization 
model (Zdziarski, Johnson \& Magdziarz 1996). This model is fast and has few 
parameters, which makes it easy to use. We would like to stress that one important 
difference between the Comptonization model and a power law is the low-energy 
cutoff. The cutoff should be present in the Comptonized spectrum around the maximum 
disc temperature if the seed photons come from the disc. The soft power law, 
extending down towards lower energies without cutoff, may significantly affect 
measurements of the disc parameters.

\begin{figure}
\begin{center}
\leavevmode
\epsfxsize=6cm \epsfbox{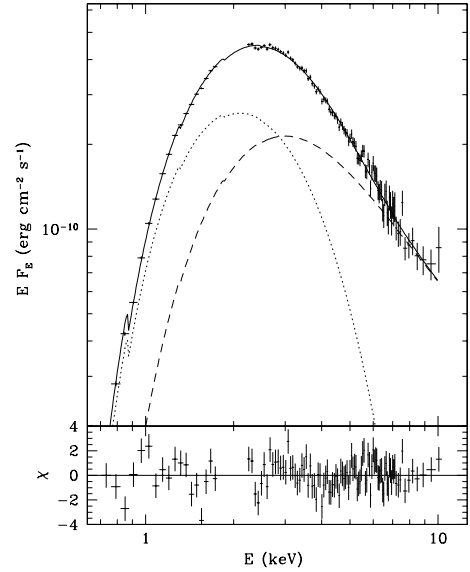}
\end{center}
\caption{The {\it ASCA}/SIS observation of LMC X-1 and a model with a non-rotating 
black hole (upper panel; see Table \ref{tab:lmc_x-1_main_fits}). The model (solid 
curve) is decomposed into the GR disc (dotted curve) and thermal Comptonization 
(dashed curve). The lower panel shows the data--model residuals divided by errors.}
\label{fig:lmc_x-1_fit}
\end{figure}

The Comptonization model has the following parameters: the electron temperature, 
$kT_e$, the asymptotic power-law photon spectral index, $\Gamma$, and the seed 
photons temperature, $kT_{\rm s}$. We have found that the electron temperature has a 
negligible effect on the fit results, so we keep it at 50 keV during our fits. We 
note, that due to lack of the data above 10 keV we cannot constrain plasma 
parameters, and we treat the Comptonization model only as a phenomenological 
component supplementary to the disc model. It is not our intention to study the hot 
plasma properties in this paper.

\begin{table*}
\centering
\caption{Fit results of the GR disc plus thermal Comptonization model to LMC X-1 {\it 
ASCA}/SIS data for five different values of the black hole spin. The inclination 
angle is $i = 50^\circ$ and the distance to the source $D$ = 50 kpc. The disc extends 
from $r_{\rm in} = r_{\rm ms}$ to $r_{\rm out} = 10^4$. The disc and Comptonization 
models are described in Sections \ref{sec:model} and \ref{sec:lmc_x-1_fits}, 
respectively. $\dot{M}_{\rm d}$ denotes accretion rate in the disc (excluding power 
dissipated in the hot phase).}
\vspace{12pt}
\begin{tabular}{lccccc}
\hline
$a$ & 0.00 & 0.25 & 0.50 & 0.75 & 0.998 \\
$N_{\rm H}$ ($10^{21}$ cm$^{-2}$) & $5.35_{-0.19}^{+0.15}$ & $5.33_{-0.18}^{+0.16}$ & 
$5.34_{-0.19}^{+0.15}$ & $5.35_{-0.19}^{+0.16}$ & $5.41_{-0.20}^{+0.16}$ \\
$\Gamma$ & $3.29_{-0.50}^{+0.37}$ & $3.30_{-0.44}^{+0.38}$ & $3.36_{-0.39}^{+0.35}$ & 
$3.43_{-0.34}^{+0.31}$ & $3.48_{-0.22}^{+0.28}$ \\
$kT_{\rm s}$ (keV) & $0.53_{-0.07}^{+0.04}$ & $0.52_{-0.06}^{+0.05}$ & 
$0.52_{-0.05}^{+0.04}$ & $0.51_{-0.04}^{+0.05}$ & $0.51_{-0.04}^{+0.03}$ \\
$M$ (M$_\odot$) & $9.7_{-1.6}^{+1.5}$ & $11.3_{-1.7}^{+2.1}$ & $14.2_{-2.3}^{+2.7}$ & 
$20.0_{-3.4}^{+3.8}$ & $45_{-8}^{+9}$ \\
$\dot{M}_{\rm d}$ ($10 ^{18}$ g s$^{-1}$) & $3.7_{-0.8}^{+0.8}$ & $3.1_{-0.7}^{+0.7}$ 
& $2.5_{-0.5}^{+0.5}$ & $1.9_{-0.4}^{+0.4}$ & $1.0_{-0.19}^{+0.12}$ \\
$\chi^2$/126 d.o.f. & 155.1 & 155.5 & 155.7 & 155.7 & 155.8 \\
\hline
\end{tabular}
\label{tab:lmc_x-1_main_fits}
\end{table*}

We fit the data by a model consisting of the GR disc, Comptonization and interstellar 
absorption. First, we assume that the disc extends down to the marginally stable 
orbit, $r_{\rm in} = r_{\rm ms}$. The fit results are presented in Table 
\ref{tab:lmc_x-1_main_fits}. We notice that goodness of the fit does not change with 
the black hole spin. However, the best fitting mass varies between $\sim$ 10M$_\odot$ 
for the non-rotating black hole to $\sim$ 40M$_\odot$ for the maximally rotating 
black hole. As we see, only the fits with $a \la 0.5$ are consistent with the upper 
mass limit, $M < 12.5$M$_\odot$ (Section \ref{sec:lmc_x-1}). 

\begin{figure}
\begin{center}
\leavevmode
\epsfxsize=9.2cm \epsfbox{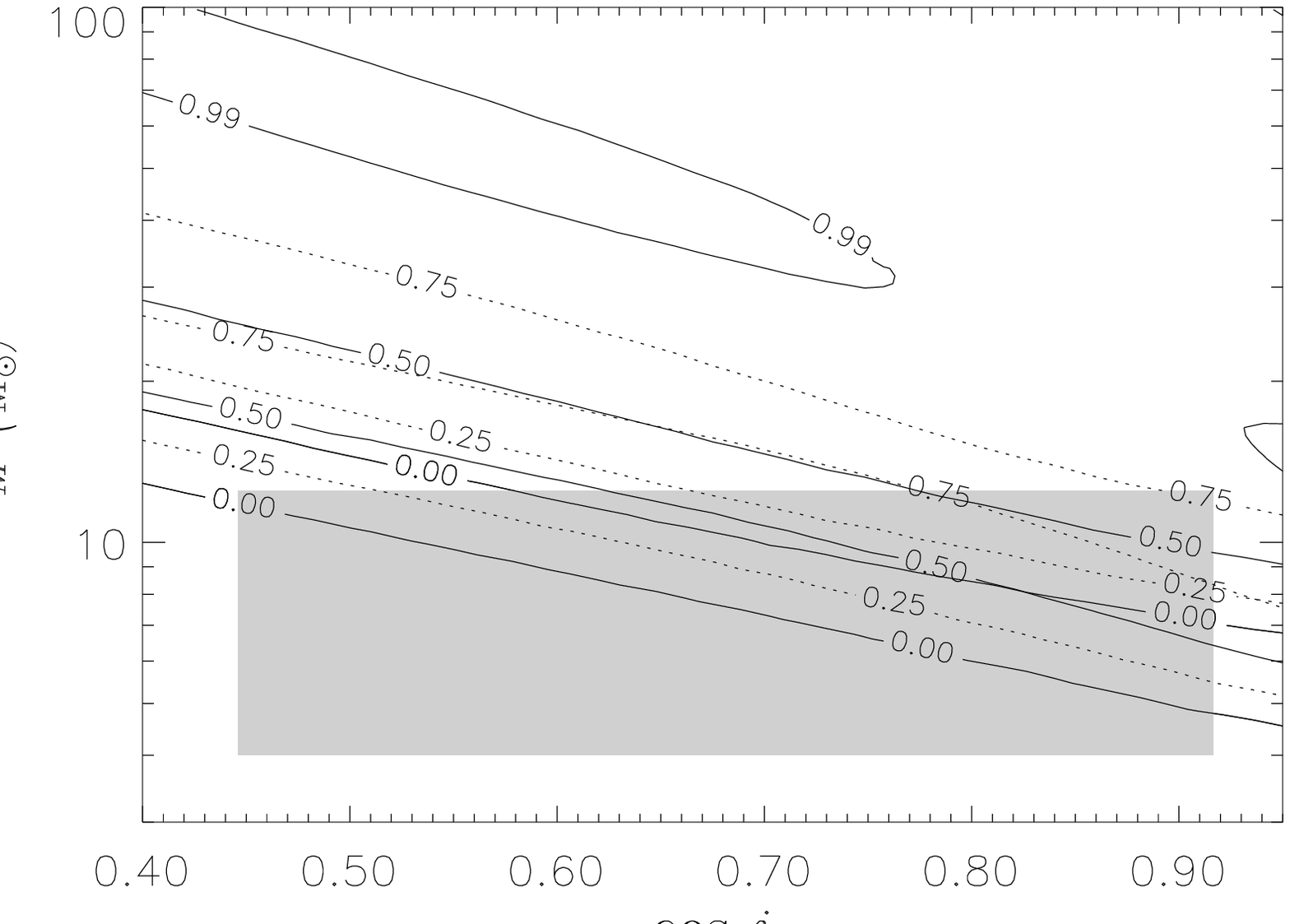}
\end{center}
\caption{Black hole mass versus cosine inclination angle for LMC X-1. Curves 
represent 90 per cent confidence contours for fits of the GR disc plus thermal 
Comptonization model. Distance of $D$ = 50 kpc has been assumed. Numbers on contour 
lines denote black hole spin. The shaded area corresponds to mass and inclination 
angle constraints from optical observations (Section \ref{sec:lmc_x-1_properties}).}
\label{fig:lmc_x-1_cont}
\end{figure}

In order to better estimate the black hole spin we fit the data in a wide range of 
inclination angles and compute 90 per cent confidence contours for inclination-mass 
relation. The results are presented on Figure \ref{fig:lmc_x-1_cont}. We find the 
fits with $a$ = 0, 0.25, 0.5 and 0.75 consistent with limits on the mass and the 
inclination angle (shaded area), and only $a = 0.998$ contour lies outside the 
acceptable area. Even if we take into account the significant uncertainty in the LMC 
X-1 distance estimation, this contour is still outside the limits. Thus, though we 
cannot precisely find the upper limit for the black hole spin, we conclude that the 
spin close to maximal is ruled-out. We stress, however, that this holds only when the 
disc extends down to the marginally stable orbit. 

If we allow for free $r_{\rm in} > r_{\rm ms}$ we cannot limit the black hole spin. 
In case of a non-rotating black hole the spectral fits do not constrain the inner 
disc radius. The best-fitting value is $r_{\rm in} \approx 100$, but with the fit 
improvement by $\Delta\chi^2 = 1.9$ only in compare to the disc extending down to 
$r_{\rm ms}$. This value of $r_{\rm in}$, however, would require a very small mass of 
the compact object, $M \approx 1.5$M$_\odot$ and a super-Eddington accretion rate, 
$\dot{M}_{\rm d} \approx 80L_{\rm Edd} c^{-2}$. From the lower mass limit, $M > 
4$M$_\odot$, we find the upper limit on the inner disc radius, $r_{\rm in} \la 30$. 
Since at this radius the black hole spin is not relevant, this limit holds for any 
spin, and we find no constraints on $a$ this time.

Within the acceptable range of $M$, $a$, $D$ and $i$ the best-fitting relative 
accretion rate in the disc, $\dot{m}_{\rm d} \equiv \dot{M}_{\rm d}c^2/L_{\rm Edd}$, 
lies between $\sim$ 1 and $\sim$ 5. In terms of the radiated power it corresponds to 
a fraction $\eta\dot{m}_{\rm d} \sim$ 0.06--0.3 of the Eddington luminosity, where 
$\eta$ is the accretion efficiency at a given spin. The total accretion rate, 
including the hard tail luminosity, is $\eta\dot{m} \sim$ 0.1--0.5. The unabsorbed 
bolometric luminosity of LMC X-1, inferred from the model, is (2.5--3.7)$\times 
10^{38}$ erg s$^{-1}$ ($3.2 \times 10^{38}$ erg s$^{-1}$ at $D$ = 50 kpc), from which 
$\sim 0.6$ is in the disc and the rest in the tail.

The parameters of the Comptonizing tail are worth a word of comment. As it can be 
seen from Table \ref{tab:lmc_x-1_main_fits}, a photon spectral index we found, 
$\Gamma \sim$ 3.3--3.5, appears to be significantly softer, than typical values of 
2.1--2.4 reported by Ebisawa et al.\ (1989) and Schlegel et al.\ (1994). However, the 
discrepancy emerges from the different model applied. When we fit our {\it ASCA\/} 
data with the simple model of the MCD and a power law, we find $\Gamma = 
2.26_{-0.16}^{+0.13}$ and $kT_{\rm in} = 0.807\pm{0.007}$ keV, which confirms that 
LMC X-1 was in the same spectral state during all these observations. A different 
result of softer spectrum was reported by Nowak et al.\ (2001), who fitted the 1996 
December 6--8 {\it RXTE\/} observation with the multicolour disc + power law model 
and found $\Gamma \approx$ 2.9--3.1 and $kT_{\rm in} \approx 0.9$ keV. 


\subsection{GRO J1655--40}
\label{sec:gro_j1655-40}

\subsubsection{Source properties}
\label{sec:gro_j1655-40_properties}

The X-ray transient GRO J1655--40 was discovered by BATSE detector on board {\it 
CGRO\/} on July 27, 1994 (Zhang et al.\ 1994). The optical counterpart was soon 
discovered by Bailyn et al.\ (1995a). The inclination of the binary system is in the 
range 63$^\circ$.7 to 70$^\circ$.7 (van der Hooft et al.\ 1998). The distance to the 
source, $D = 3.2\pm0.2$ kpc was derived from jet kinematics (Hjellming \& Rupen 
1995). The mass function was first obtained by Bailyn et al.\ (1995b) and then 
improved by Orosz \& Bailyn (1997), who found $f_M = 3.24\pm0.09$M$_\odot$ and 
classified the companion star as F3 {\sc iv}--F6 {\sc iv}. In the same paper Orosz \& 
Bailyn determined the mass of the compact object with unprecedented accuracy, finding 
$M_X = 7.02\pm0.22$M$_\odot$. However, Shahbaz et al.\ (1999) pointed out that in 
calculating the radial velocity semi-amplitude, Orosz \& Bailyn (1997) used 
observations both during X-ray quiescence and during an X-ray outburst, which could 
lead to an incorrect result. Shahbaz et al.\ (1999) used only the X-ray quiescence 
data, and found different value of the mass function, $f_M = 2.73\pm0.09$M$_\odot$, 
the mass ratio, $q$ = 0.337--0.436, and the compact object mass, $M$ = 
5.5--7.9M$_\odot$ (95 per cent confidence). With this mass, GRO J1655--40 is one of 
the most firmly established black hole candidates. 

Several authors have tried to estimate the black hole spin in GRO J1655--40, however 
they came up with different and conflicting conclusions. Zhang et al.\ (1997a) 
fitted the August 1995 {\it ASCA\/} data of the source with the MCD
model with relativistic corrections. They found the inner disc radius of 23 km for a 
7M$_\odot$ black hole, significantly smaller than the marginally stable orbit radius 
in the Schwarzschild metric, $R_{\rm ms} = 62$ km. Their conclusion is that GRO 
J1655--40 contains a Kerr black hole rotating at 0.7--1.0 of the maximum rate. 
Sobczak et al.\ (1999) applied a similar approach to soft state {\it RXTE\/} 
observations, finding average inner disc radius of $R_{\rm in} = 4.2R_g$. They 
associated this radius with the marginally stable orbit corresponding to the spin of 
$a = 0.5$. Taking into account uncertainties in the spectral hardening factor, 
$f_{\rm col}$, and the distance to the source, they concluded that $a < 0.7$. 
Makishima et al.\ (2000) applied the relativistic accretion disc model in 
Schwarzschild metric to the same data and found that if the black hole is not 
rotating, its mass, $M = 2.9\pm0.1$M$_\odot$, is incompatible with optical 
constraints, therefore there must be a rotating black hole in GRO J1655--40.

An alternative approach to estimating the black hole spin is analysis of 
quasi-periodic oscillations (QPOs) observed in the power spectra. Remillard et al.\ 
(1999) analysed {\it RXTE\/} observations of GRO J1655--40 and found four 
characteristic QPOs. Three of them occupy relatively stable frequencies of about 0.1, 
9 and 300 Hz. Cui, Zhang \& Chen (1997) associated the highest frequency QPO with the 
nodal precession frequency of the tilted disc and derived $a = 0.95$. On the other 
hand, Stella, Vietri \& Morsink (1999) interpreted the highest frequency QPO in terms 
of the periastron precession frequency, and the $\sim$ 9 Hz QPO as a second harmonic 
of the nodal precession frequency. This interpretation yielded $a \sim 0.1$. Gruzinov 
(1999) linked the $\sim$ 300 Hz QPO with the emission of the bright spot at the 
radius of the maximal proper radiation flux, and inferred $a \sim 0.6$.

In X-rays GRO J1655--40 is highly variable and undergoes spectral
transitions similar to that of classical X-ray novae. Ueda et al.\ 
(1998) analysed the four {\it ASCA\/} observations between August 1994
and March 1996. They distinguished four distinct states, named
``high", ``low", ``dip" and ``off". The high state was observed by
BATSE during the outburst in the energy range of 20--100 keV (Tavani
et al.\ 1996; Zhang et al.\ 1997b), while the low state corresponds to
times when the source was weak also in the BATSE range. This high
state can be associated with the soft state, since its spectrum is
dominated by the ultra-soft disc component with additional high-energy
power law. Zhang et al.\ (1997b) found the power-law photon spectral
index of $\Gamma = 2.43\pm0.3$ from the BATSE data simultaneous to the
{\it ASCA\/} observation analysed in this paper. A long-time {\it
  RXTE\/} monitoring (Sobczak et al.\ 1999) shows that the soft state
can be described by the MCD model with $kT_{\rm in} \sim 0.7$ keV plus
a power law with photon spectral index of $\Gamma \sim$ 2--3.

\subsubsection{Spectral fits}
\label{sec:gro_j1655-40_fits}

GRO J1655--40 was observed by {\it ASCA\/} on August 15--16, 1995. This is the 
observation from epoch III from Ueda et al.\ (1998), who associated it the high X-ray 
state of the source. From this observation we have extracted the GIS spectrum with 
the net exposure of 3810 s after dead-time correction. We use the data in the 1.0--10 
keV range. 

We assume the distance to the source, $D = 3.2$ kpc (Hjellming \& Rupen 1995), and 
the inclination angle of $70^\circ$ (Orosz \& Bailyn 1997; van der Hooft et al.\ 
1998), unless stated otherwise. We do not impose any constraints on the black hole 
mass during the fits, but check its consistency with the limits obtained from the 
optical observations, $5.5$M$_\odot \leq M \leq 7.9$M$_\odot$ (Shahbaz et al.\ 1999; 
see Section \ref{sec:gro_j1655-40}), {\it a posteriori}.

\begin{figure}
\begin{center}
\leavevmode
\epsfysize=8.4cm \epsfbox{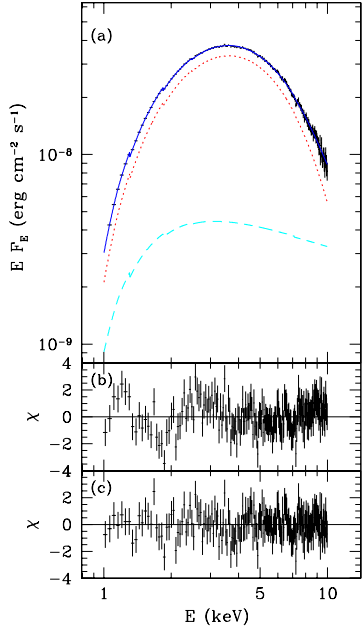}
\end{center}
\caption{The {\it ASCA}/GIS spectrum of GRO J1655--40 and the best-fitting model with 
$a = 0.75$ (see Table \ref{tab:gro_j1655_main_fits}). Panel $(a)$ shows the data and 
the model. The model (solid curve) is decomposed into the GR disc (dotted curve) and 
the power law (dashed curve). The residuals (data--model divided by errors) shown in 
panel $(c)$ correspond to this fit. For comparison, on panel $(b)$ we present the 
residuals of the fit without absorption edge.}
\label{fig:gro_j1655_fit}
\end{figure}

The high-energy tail beyond the disc spectrum is less significant than in the case of 
LMC X-1 (see Figures \ref{fig:lmc_x-1_fit} and \ref{fig:gro_j1655_fit}), and we find 
that the particular choice between a Comptonization model and a power law does not 
affect the disc fit results, so we chose a power law with the photon spectral index 
fixed at $\Gamma = 2.4$ (Zhang et al.\ 1997b). On the other hand, the tail cannot be 
neglected. A fit without the tail is worse by $\Delta\chi^2 = +90$ (at 188 d.o.f.) in 
compare to our best fit (see below) and shows increasing positive residuals towards 
higher end of the energy band.

We fit the data by a model consisting of the GR disc, power law and interstellar 
absorption. First we assume the disc extends down to the marginally stable orbit, 
$r_{\rm in} = r_{\rm ms}$. The model fits the data fairly well, with $\chi^2 \sim 
230/187$, however there is a strong residual pattern below $\sim$ 3 keV (see panel 
$(b)$ on Figure \ref{fig:gro_j1655_fit}). The pattern is likely to be generated by 
atomic processes. However, the interstellar absorption, both on neutral and ionized 
matter, cannot explain this pattern. Varying elemental abundances in the absorber 
does not improve the fit. Also, the pattern cannot be fitted by ionized Compton 
reflection. We notice that relativistically broadened edge is required to explain the 
observed spectrum. 

Therefore, we included a smeared absorption edge ({\tt smedge} model in {\sc xspec}; 
Ebisawa 1991) in our model spectrum. {\tt smedge} is only a phenomenological formula 
invented to reproduce the relativistic smearing and its parameters should not be 
taken as real physical quantities. In particular, when the smearing is introduced the 
model edge threshold energy is {\em lower} than the rest frame threshold energy of 
the physical edge. We have found that inclusion of the absorption edge improves the 
fit considerably, yielding much better $\chi^2 \sim$ 165/185. The smearing width of 
the edge is poorly constrained around $\sim 0.5$ keV, so we fix it at this value 
during all fits. The threshold energy, typically $1.25\pm0.05$ keV, might be 
consistent with the Ne-like Fe L-shell absorption edge (at 1.26 keV), originating in 
the ionized disc. Detailed analysis of this feature is however beyond the scope of 
this paper.

Finally, our best-fitting model consists of the GR disc, the power law, the 
interstellar absorption and the absorption edge. Fit results for five values of the 
black hole spin are presented in Table \ref{tab:gro_j1655_main_fits}. The data and 
the best-fitting model for $a = 0.75$ are shown on Figure \ref{fig:gro_j1655_fit}.

\begin{table*}
\centering
\caption{Fit results of the GR disc plus power law and an absorption edge model to 
GRO J1655--40 {\it ASCA}/GIS data for five different values of the black hole spin. 
The inclination angle is $i = 70^\circ$ and the distance to the source $D$ = 3.2 kpc. 
The power-law photon index is fixed at $\Gamma = 2.4$. The disc extends from $r_{\rm 
in} = r_{\rm ms}$ to $r_{\rm out} = 10^4$. The disc model is described in Section 
\ref{sec:model}.} 
\vspace{12pt}
\begin{tabular}{lccccc}
\hline
$a$ & 0.00 & 0.25 & 0.50 & 0.75 & 0.998 \\
$N_{\rm H}$ ($10^{21}$ cm$^{-2}$) & $7.48_{-0.18}^{+0.17}$ & $7.48_{-0.17}^{+0.17}$ & 
$7.50_{-0.18}^{+0.17}$ & $7.58_{-0.18}^{+0.16}$ & $7.70_{-0.17}^{+0.15}$ \\
$E_{\rm edge}$ (keV) & $1.25_{-0.05}^{+0.05}$ & $1.25_{-0.05}^{+0.05}$ & 
$1.25_{-0.05}^{+0.05}$ & $1.25_{-0.05}^{+0.04}$ & $1.23_{-0.04}^{+0.04}$ \\
$\tau_{\rm edge}$ & $0.13_{-0.04}^{+0.03}$ & $0.13_{-0.03}^{+0.04}$ & 
$0.14_{-0.03}^{+0.04}$ & $0.16_{-0.03}^{+0.04}$ & $0.17_{-0.03}^{+0.03}$ \\
$M$ (M$_\odot$) & $2.65_{-0.02}^{+0.03}$ & $3.19_{-0.03}^{+0.03}$ & 
$4.07_{-0.04}^{+0.04}$ & $5.84_{-0.05}^{+0.06}$ & $16.2_{-0.2}^{+0.1}$ \\
$\dot{M}_{\rm d}$ ($10 ^{18}$ g s$^{-1}$) & 3.21$_{-0.03}^{+0.04}$ & 
2.71$_{-0.04}^{+0.04}$ & 2.15$_{-0.02}^{+0.03}$ & 1.51$_{-0.02}^{+0.02}$ & 
0.560$_{-0.007}^{+0.007}$ \\
$\chi^2$/185 d.o.f. & 166.1 & 165.3 & 164.3 & 163.4 & 165.0 \\
\hline
\end{tabular}
\label{tab:gro_j1655_main_fits}
\end{table*}

\begin{table*}
\centering
\caption{The same as in Table \ref{tab:gro_j1655_main_fits}, but now $r_{\rm in}$ is 
a free fit parameter. For the sake of clarity, we show only $r_{\rm in}$ and $M$.} 
\vspace{12pt}
\begin{tabular}{lccccc}
\hline
$a$ & 0.00 & 0.25 & 0.50 & 0.75 & 0.998 \\
$r_{\rm in}$ & $7.6_{-1.6}^{+12}$ & $6.6_{-1.4}^{+9.0}$ & $5.5_{-1.3}^{+6.1}$ & 
$3.8_{-0.6}^{+4.7}$ & $3.7_{-2.5}^{+4.4}$\\
$M$ (M$_\odot$) & $2.7_{-1.1}^{+0.0}$ & $3.2_{-0.4}^{+0.0}$ & $4.0_{-1.4}^{+0.1}$ & 
$5.9_{-1.9}^{+0.0}$ & $10_{-6.4}^{+6.3}$ \\
$\chi^2$/184 d.o.f. & 165.7 & 165.0 & 164.1 & 163.2 & 163.7 \\
\hline
\end{tabular}
\label{tab:gro_j1655_rin_fits}
\end{table*}

\begin{figure}
\begin{center}
\leavevmode
\epsfxsize=9.2cm \epsfbox{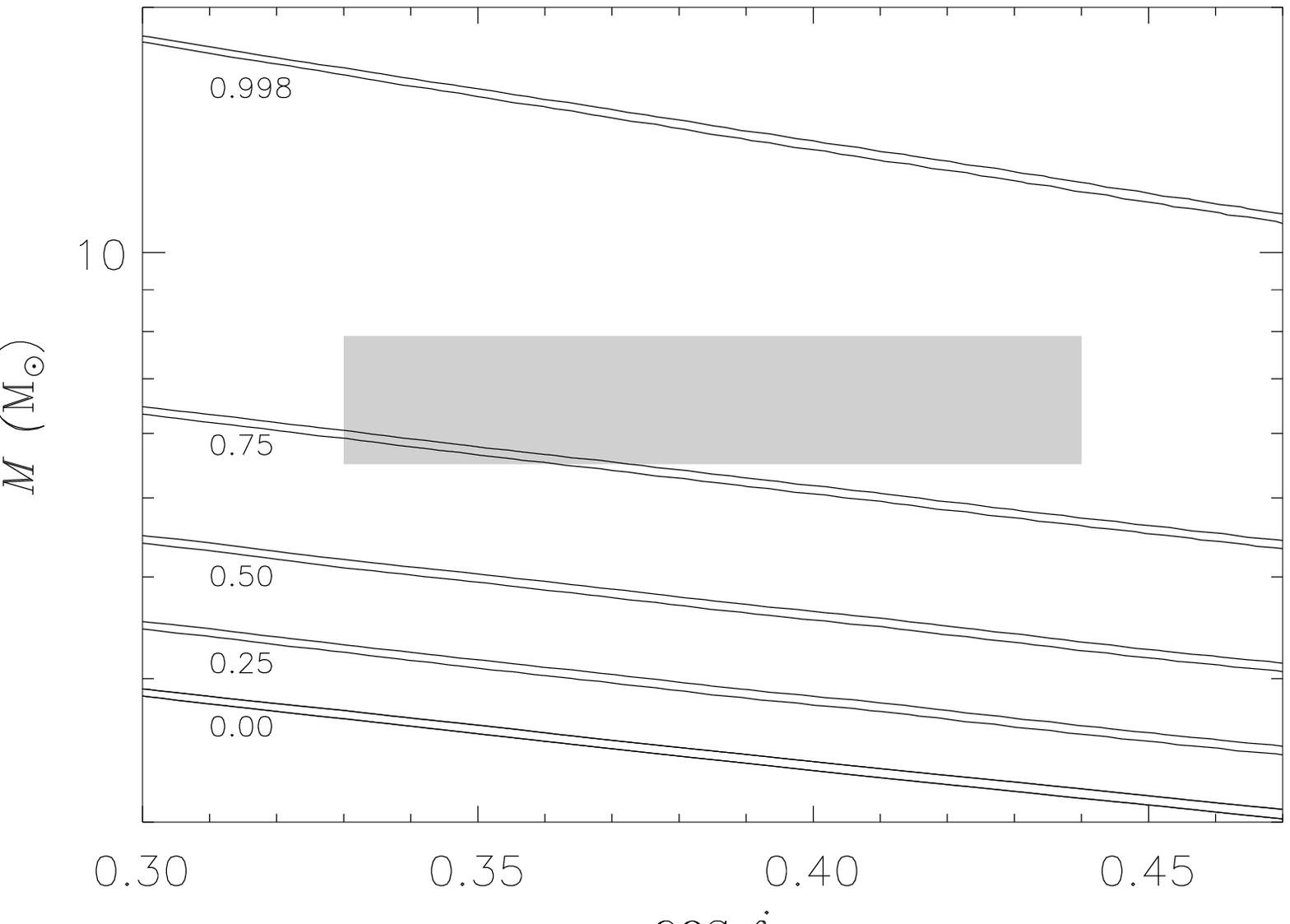}
\end{center}
\caption{Black hole mass versus cosine inclination angle for GRO J1655--40. Curves 
represent 90 per cent confidence contours for fits of the GR disc plus power law 
model. Distance of $D$ = 3.2 kpc has been assumed. Numbers below contour lines denote 
the black hole spin. The shaded area corresponds to mass and inclination angle 
constraints from optical observations (Section \ref{sec:gro_j1655-40_properties}).}
\label{fig:gro_j1655-40_cont}
\end{figure}

As it was in the case of LMC X-1, spectral fits do not constrain the
black hole spin.  However, the spin can be constrained taking into
account mass limitations. Only the fit with $a = 0.75$, yielding $M =
5.84_{-0.05}^{+0.06}$M$_\odot$, is consistent with the mass constraint
of (5.5--7.9)M$_\odot$, inferred from the optical observations by
Shahbaz et al.\ (1999). Again, we fit the data in a range of
inclination angles and compute 90 per cent confidential contours for
inclination-mass relation. The result is presented on Figure
\ref{fig:gro_j1655-40_cont}. Due to better quality of the data and due
to the fact that GRO J1655--40 spectrum is dominated by the disc
emission, these contours are much narrower than in the case of LMC
X-1. Only the $a = 0.75$ contour overlaps with the shaded area
indicating allowable values of the inclination angle and the mass.
More precise limits on the spin can be found by interpolating between
the values found from the fits. If the GR effects are neglected, for
the given spectrum the disc must keep constant inner radius while
varying $M$ and $a$.  Providing that $r_{\rm in} = r_{\rm ms}$, this
leads to a simple relation between the mass and the spin $M \propto
r_{\rm ms}^{-1}(a)$, where $r_{\rm ms}(a)$ is given by equation
(\ref{eq:rms}). The GR effects modify this relation and we have found
that a function $M = C_1 + C_2 r_{\rm ms}^{-1}(a)$ approximates the
data well, so we use it for interpolation. Taking into account
uncertainties in the distance to the source, we have found that for
given limits $5.5 \leq M/$M$_\odot \leq 7.9$ and $64^\circ \leq i \leq
71^\circ$ the black hole spin is $0.68 \leq a \leq 0.88$. In this
range of spin values returning radiation is negligible (see
Section \ref{sec:model}), and does not affect our result.

Next, we fit the data allowing the inner disc radius, $r_{\rm in}$, to be a free 
parameter. The results are presented in Table \ref{tab:gro_j1655_rin_fits}. In all 
fits we find that the inner disc radius is consistent with $r_{\rm ms}$, however 
larger values are possible. Now the $a = 0.998$ fit became acceptable with $M = 
10_{-6.4}^{+6.3}$M$_\odot$. Since the upper limits for the mass remained virtually 
the same after freeing $r_{\rm in}$ (see Tables \ref{tab:gro_j1655_main_fits} and 
\ref{tab:gro_j1655_rin_fits}), the lower limit on the spin also remained the same, $a 
\geq 0.68$, this time.

Within the acceptable range of $M$, $D$ and $i$ (assuming $r_{\rm in} = r_{\rm ms}$) 
the relative accretion rate, $\dot{m}_{\rm d}$, is between $\sim$ 0.6 and $\sim$ 1.9. 
In terms of the radiated power it corresponds to a fraction $\eta\dot{m}_{\rm d} 
\sim$ 0.1--0.2 of the Eddington luminosity. The unabsorbed bolometric luminosity of 
the disc in GRO J1655--40, computed from the model, is $(1.2$--$1.6)\times 10^{38}$ 
erg s$^{-1}$ ($1.4 \times 10^{38}$ erg s$^{-1}$ at $D = 3.2$ kpc). In the observed 
energy range only a small fraction of the luminosity is radiated in the tail.


\section{Discussion and conclusions}
\label{sec:discussion}

\subsection{Black hole mass and spin}

Optical and UV observations of black hole binaries can provide with the mass 
function, which in turn can help estimating, to some extent, the black hole mass and 
the inclination angle of the binary. We know about a dozen of X-ray Galactic binaries 
for which the mass function implies the compact object mass larger than 3M$_\odot$, 
making them good candidates for black holes. However, the black hole spin cannot be 
estimated this way.

In this paper we have shown an application of the GR disc model to the soft state 
X-ray spectra of LMC X-1 and GRO J1655--40 and demonstrated how it can restrict the 
black hole spin. The model depends on five basic parameters: the black hole mass and 
spin, the disc inclination angle, the accretion rate and the inner disc radius. 
Generally, the spectral shape changes only slightly with these parameters, and only 
two out of them can be established uniquely by fitting the observational data. 
Additional assumptions about some of the parameters are necessary. In most of our 
fits we have fixed $i$ and assumed $r_{\rm in} = r_{\rm ms}$. Then, for several fixed 
values of $a$ we have successfully fitted $M$ and $\dot{M}$. Results presented in 
Tables \ref{tab:lmc_x-1_main_fits} and \ref{tab:gro_j1655_main_fits} show that 
$\chi^2$ remained virtually the same through the whole range of $a$ and the black 
hole spin cannot be reckoned this way. On the other hand, at a given spectral shape 
there is strong correlation between the parameters, in particular between $M$ and $a$ 
when $i$ and $r_{\rm in}$ are fixed. With an independent estimation of the mass and 
the inclination angle, we can constrain the black hole spin. The tighter constraints 
on the mass and the inclination, the more accurate spin estimation can be obtained.

The mass and the inclination angle of LMC X-1 are evaluated rather
poorly. Therefore, it is not possible to give a good constrain on the
black hole spin. We can only rule out a black hole which is close to,
or maximally rotating. GRO J1655--40 gave us better chance. The mass
and the inclination angle are measured with relatively high accuracy.
Also, due to better quality of the X-ray data, we have obtained
smaller statistical errors on $M$. As a result, we could limit the
spin, $0.68 \leq a \leq 0.88$ (when $r_{\rm in} = r_{\rm ms}$). This
result is only weakly affected by neglecting returning photons, which
are negligible for spin less then 0.9.

We underline, that when we allow for free $r_{\rm in} > r_{\rm ms}$, no upper limit 
for the black hole spin can be imposed. This is because for a given spectral shape 
increase of $r_{\rm in}$ with fixed $M$ and $i$ leads to increase of $a$. However, 
there are clues that the accretion disc in BHB extends down to the marginally stable 
orbit all the time in the soft (high or very high) state.  

Long-term 1996-1997 {\it RXTE\/} monitoring of GRO J1655--40 (Sobczak et al.\ 1999) 
shows remarkable constancy of the MCD inner disc radius over the period of a few 
hundred days, while the source went through the high and very high spectral states. 
There is an exception of a few observations during the very high state, where the 
inner radius dropped suddenly by factor $\sim 3$. Merloni, Fabian \& Ross (2000) 
interpret it as a drop in the disc accretion rate, which caused increase of the 
hardening factor and decrease of the {\em apparent\/} inner radius, while the actual 
$r_{\rm in}$ has not changed. If the inner disc radius remained constant indeed, it 
is reasonable to assume that the disc extended down to the marginally stable orbit 
all the time when the source was in the high or very high state, including also the 
{\it ASCA\/} observation analysed in this paper. 

Long-term 1996-1998 {\it RXTE\/} monitoring of LMC X-1 (Wilms et al.\ 2001) shows in 
turn significant variations of the MCD inner disc radius. However, it is not clear 
whether these variations can be attributed to real changes of $r_{\rm in}$. They 
might be, at least in part, due to systematic errors in the model. 

So, does $r_{\rm in}$ vary during the soft state or not?  If the cold
disc in either of the above observations was indeed truncated, it
should turn into an optically thin inner flow below $r_{\rm in}$. Hot
plasma in the inner flow could be a source of the high-energy tail
beyond the disc spectrum. However, optically thin solution can exist
only for luminosities lower than a few per cent of the Eddington
luminosity. At higher accretion rates (and higher densities) Coulomb
transfer of energy from the protons to the electrons becomes
efficient, the optically thin flow becomes radiatively efficient and
collapses to the optically thick Shakura-Sunyaev disc (Chen et al.\ 
1995; Esin, McClintock \& Narayan 1997). In both observations analysed
in this paper luminosity exceeded 10 per cent of $L_{\rm Edd}$,
therefore we should not expect any hot, optically thin flow. Instead,
we propose that the cold disc extends all the way down to the
marginally stable orbit, and the high-energy tail emission is produced
in active regions above the disc. Hence, we find the assumption of
$r_{\rm in} = r_{\rm ms}$ as well founded, and our best constraint on
the black hole spin in GRO J1655--40 is $0.68 \leq a \leq 0.88$ and in
LMC X-1 we rule a black hole rotating with the spin close to maximum.

The most spectacular feature of GRO J1655--40 is its radio jets (Tingay et al.\ 1995; 
Hjellming \& Rupen 1995), observed also in another high-spin black hole candidate, 
GRS 1915+105 (Mirabel \& Rodr\'{\i}guez 1994). Co-existence of relativistic jets and 
rotating black holes in BHB makes and interesting link to the problem of radio 
dichotomy of quasars, and supports the black hole ``spin paradigm", according to 
which jets in quasars are powered by rotating black holes (e.g.\ Moderski, Sikora \& 
Lasota 1998). Since no strong radio jets were found in LMC X-1, the non-rotating 
black hole should be expected in this system. Unfortunately, our results do not allow 
us to make such claims. Further systematic studies of several jet and non-jet sources 
could provide with more evidences for the spin paradigm.

\subsection{Compton reflection}

The X-ray spectra of LMC X-1 and GRO J1655--40 are similar to the soft state spectrum 
of Cyg X-1 (Gierli{\'n}ski et al.\ 1999), though the first two sources are 
significantly brighter. An important difference is that we do not find Compton 
reflection features here, while Cyg X-1 in the soft state showed strong reflection 
component with covering angle $\Omega/2\pi \sim 0.7 $ and a broad Fe K$\alpha$ line. 
This can be explained by the instrumental limitations or/and by the intrinsic 
difference in the source geometry and energetics. The Compton reflection from the 
cold matter can be identified by Fe K$\alpha$ features around 7 keV and by reflection 
continuum, which peaks around 30 keV in the $E F_E$ spectrum. Since {\it ASCA\/} can 
observe only up to $\sim$ 10 keV, the detection of the reflection continuum with {\it 
ASCA\/} data only might be difficult, if not impossible. The iron K edge and line 
parameters are sensitive to the shape of the underlying continuum, which cannot be 
properly established without the high-energy data. For example, Dotani et al.\ (1997) 
analysed the {\it ASCA\/} data of Cyg X-1 and did not find any Fe line. 
Gierli{\'n}ski et al.\ (1999) re-analysed the same data jointly with the simultaneous 
{\it RXTE\/} observation, and found the presence of the iron line with high 
statistical significance. LMC X-1 spectrum has poor statistics above $\sim$ 5 keV, so 
the reflection features can escape the detection. The high-energy tail in GRO 
J1655--40 is much weaker than in Cyg X-1. At 7 keV the disc radiates $\sim 3.5$ times 
more energy than the tail. Therefore, the Fe K-shell features will be very weak in 
the total spectrum, and difficult to detect. In the soft state of Cyg X-1 the 
situation is opposite: the disc emission at 7 keV is negligible (see Figure 8 in 
Gierli{\'n}ski et al.\ 1999 and Figure \ref{fig:lmc_x-1_fit} in this paper). 
Moreover, the Fe K-shell features coming from the disc around the fast-spinning black 
hole are significantly smeared, which additionally hinders the detection.

We note, that {\it ASCA\/} and {\it RXTE\/} detected characteristic
iron features in GRO J1655--40, at different periods (Ueda et al.\ 
1998; Tomsick et al.\ 1999; Ba{\l}uci{\'n}ska-Church \& Church 2000;
Yamaoka et al.\ 2000).  However, precise {\it ASCA}/{\it RXTE\/}
observations show that these features can be consistently explained by
resonant absorption lines in the corona above the disc {\em without}
reflection components (Yamaoka et al.\ 2000).

\subsection{Disc stability}

An unclear issue is stability of the cold disc. Both sources are very bright with 
luminosities being a significant fraction of the Eddington luminosity ($\dot{m}_{\rm 
d} \sim$ 1--5 is a very conservative limit for LMC X-1). A radiation pressure 
dominated region, which is thought to be unstable, arises in the disc above the 
critical accretion rate, $\dot{m}_{\rm crit}$. In the pseudo-Newtonian approximation, 
for a 10M$_\odot$ black hole and $\alpha = 0.1$, $\dot{m}_{\rm crit} \approx 0.64$ 
(Gierli{\'n}ski et al.\ 1999; taking into account accretion in the cold disc only, 
i.e.\ $f = 0$). Therefore, the discs in both sources seem to be dominated by the 
radiation pressure. Since we consider here rotating black holes, we checked this 
result by calculating the ratio of the radiation pressure to the gas pressure in the 
Kerr metric (Novikov \& Thorne 1972), for all acceptable fit parameters in Tables 
\ref{tab:lmc_x-1_main_fits} and \ref{tab:gro_j1655_main_fits}. We found that in both 
sources the inner part of the cold accretion disc below several tens of $R_g$ is 
dominated by the radiation pressure indeed. Thus, both discs should be unstable 
against secular and thermal instabilities. However, despite this inconvenience the 
discs in LMC X-1 and GRO J1655--40 apparently {\em do} exist. 

\begin{figure}
\begin{center}
\leavevmode
\epsfysize=7cm \epsfbox{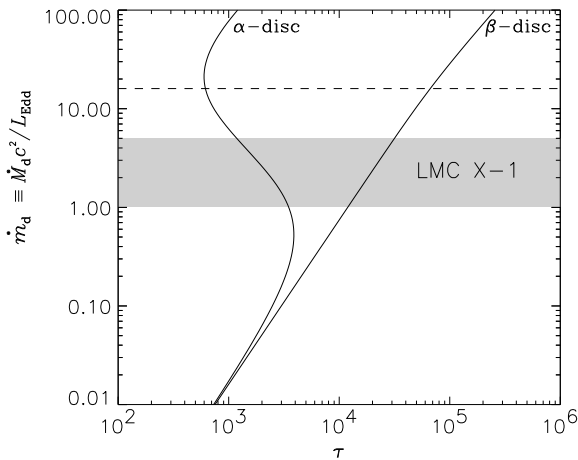}
\end{center}
\caption{Two solutions of the optically thick disc around a 10M$_\odot$ non-rotating 
black hole at the radius of the most significant instability of $r$ = 15.1. The 
viscosity parameter $\alpha$ is 0.1. The left-hand curve represents the solution with 
$T_{r\phi} \propto  P_{\rm rad} + P_{\rm gas}$ ($\alpha$-disc). The right-hand curve 
shows the solution with $T_{r\phi} \propto P_{\rm gas}$ ($\beta$-disc). The shaded 
region shows the most conservative limits on the disc accretion rate of LMC X-1 at $a 
= 0$. The dashed line represents the Eddington luminosity (efficiency $\eta = 1/16$). 
The inner disc is dominated by the radiation pressure, where the $\alpha$-disc 
solution is unstable and the $\beta$-disc solution is stable.}
\label{fig:alpha_beta}
\end{figure}

This riddle probably arises from our lack of understanding of microscopic mechanisms 
of accretion. The instability develops in the standard $\alpha$-prescription theory, 
where $r\phi$-element of the viscous stress tensor, $T_{r\phi}$, is proportional to 
the total pressure, i.e.\ to the sum of the gas pressure and the radiation pressure. 
The viscosity parameter $\alpha$ includes not well understood physics of viscosity, 
which is supposed to be due to chaotic magnetic fields and turbulence in the gas 
flow. It is not quite clear why $T_{r\phi}$ should be proportional to the total 
pressure. If we assume that $T_{r\phi}$ is proportional to the gas pressure only (the 
so-called $\beta$-disc; see e.g.\ Stella \& Rosner 1984), the disc becomes stable in 
the radiation pressure dominated region (Figure \ref{fig:alpha_beta}). The existence 
of the radiation pressure dominated discs should impose significant constraints on 
the theoretical models of viscosity in the accretion flow.

\subsection{Model reliability}

One important question we should ask is reliability of the model and parameters 
inferred. While modelling the disc spectrum in the ``proper" way one should solve the 
following four problems. (i) The radial disc structure, (ii) the vertical disc 
structure, (iii) the radiation transfer and (iv) the relativistic effects. Our 
understanding of the first three problems is poor and there are still a lot of open 
questions concerning the viscosity prescription, vertical distribution of 
gravitational energy dissipation, effects of radiation pressure and irradiation of 
the disc, importance of bound-free opacity and heavy elements, and importance of 
Comptonization, just to name the most important issues. Consequently, every single 
prediction available in the literature is based on simplifying assumptions and is 
model dependent. On the other hand, the simplest solution, i.e.\ the sum of 
blackbodies gives the best description of the observational data. For example, a 
smooth multicolour spectrum with no ionization edges consistently reproduces the AGN 
observations in which the Lyman edge (expected by more advanced models) is not 
observed. As we have shown in this chapter, the multicolour model fits the black hole 
candidate spectra perfectly as well.

We correct the emerging spectra for two effects that significantly affect 
measurements of the disc normalization and temperature. The first one is a deep 
gravitational potential in the vicinity of a fast rotating black hole. All special 
and general relativistic effects are treated to high accuracy by means of the 
transfer function, as was described in Section \ref{sec:model}. The second important 
aspect is the effect of electron scattering, which we approximate by a single colour 
temperature correction factor, $f_{\rm col}$, using a diluted blackbody as a local 
spectrum (see Section \ref{sec:model}). Under certain simplifying assumptions Shimura 
\& Takahara (1995) computed the vertical disc structure around a Schwarzschild 
stellar-mass black hole and solved the radiation transfer. They found that for the 
viscosity parameter $\alpha = 0.1$ the local spectrum can be represented by a diluted 
blackbody when Comptonization is effective, which takes place in the inner disc 
region when $\dot{m}_{\rm d} \ga 1$. For $M = 10$M$_\odot$ and $\dot{m}_{\rm d} = 1$ 
they found that the sum of diluted blackbodies fits the whole disc spectrum very well 
with the best-fitting hardening factor of $f_{\rm col} = 1.7$ (see Figure 2 in 
Shimura \& Takahra 1995). Merloni et al.\ (2000) made a similar comparison between 
the multicolour disc and a more realistic disc model with the vertical temperature 
structure and radiation transfer solved and came up with a comparable result. The 
multicolour disc spectra fit the realistic disc spectra well, and for $\dot{m}_{\rm 
d} \ga 1$ the hardening factor is $f_{\rm col} \approx 1.8$. In both observations in 
this paper $\dot{m}_{\rm d} \sim 1$, therefore, within the trustworthiness limits of 
the Shimura \& Takahara and Merloni et al.\ calculations, we find our model reliable. 
A small variation in the hardening factor does not change our results significantly. 
For example, the GRO J1655--40 spin for $f_{\rm col} = 1.8$ is $0.60 \leq a \leq 
0.83$.

\section*{Acknowledgements}

We thank Andrzej A. Zdziarski and Chris Done for discussions and
valuable comments.  This research has been supported in part by the
Polish KBN grants 2P03D00514, 2P03D00614, 2P03C00619p0(1,2) and a
grant from the Foundation for Polish Science.


\appendix
\section{Construction of the photon transfer function}
\label{app:transfer_function}

We consider a gravitational field of a rotating black hole with the mass $M$, and the 
angular momentum, $J$. In both appendices we use $G = c = 1$ units, the $(-+++)$ 
signature and the Boyer-Lindquist coordinates, $(t, R, \theta, \phi)$. Non-zero 
components of the metric tensor are given by
\begin{eqnarray}
\gtt & = & -(1-2r/\Sigma), \nonumber \\
\gtp & = & -2ar\sin^2\theta/\Sigma, \nonumber \\
\gpp & = & \left( r^2+a^2+{2a^2r\sin^2\theta \over \Sigma} \right)\sin^2 \theta, 
\nonumber \\
g_{rr} & = & \Sigma/\Delta, \nonumber \\
g_{\theta \theta} & = & \Sigma,
\end{eqnarray}   
where
\begin{eqnarray}
\lefteqn{\Delta = r^2+a^2-2r,} \nonumber \\
\lefteqn{\Sigma = r^2+a^2\cos^2\theta,} 
\end{eqnarray}
and we use dimensionless distance and specific angular momentum parameter
\begin{equation}
r={R \over M},~~~~a={J \over M^2}.
\end{equation}

Our approach to calculation of the disc photon trajectories mostly follows that of 
Cunningham (1975). A null geodesic in the Kerr metric is described by two constants 
of motion, $\lambda$ and $\xi$, defined as
\begin{equation}
\lambda={L \over EM},~~~~\xi^2={C \over E^2M^2}, 
\end{equation}
where $E$ is the photon energy at infinity, $L$ is projection of the photon angular 
momentum on the symmetry axis and $C$ is the Carter constant (in the definition of 
$\xi$ we took into account the condition $C \ge 0$ satisfied by trajectories 
intersecting equatorial plane.)

We take an usual assumption that the radial velocity in the disc can be neglected and 
its motion is Keplerian. In the non-rotating frame, the angular velocity and the 
linear velocity, respectively, are given by
\begin{equation}
\Omega_{\rm K}= {1 \over M} {1 \over r^{3/2}+a},
\end{equation}
and
\begin{equation}    
V_{\rm K}={A \over r^2 \sqrt{\Delta}} \left( \omega_{\rm K}-{2 a r
\over A} \right),
\end{equation}
where
\begin{equation}
A=r^4+a^2r^2+2a^2r,
\end{equation}
and $\omega_{\rm K}=M \Omega_{\rm K}$. Then, we get the following relation between 
the constants of motion and the emission angles in the disc rest frame (e.g., 
Cunningham \& Bardeen 1973)
\begin{eqnarray}
\lambda & = & {\sin\theta_e \sin \pem+ V_{\rm K} \over (r_e^2 \Delta^{1/2}+
2 a r_e V_{\rm K})/ A+\omega_{\rm K} \sin\theta_e \sin \pem},
 \nonumber \\
\xi & = & \left( {A \over \Delta } \right)^{1/2} (1-
V_{\rm K}^2)^{-1/2}(1-\lambda \omega_{\rm K}) \ct.
\label{const}
\end{eqnarray}
where $r_e$ is the initial photon radius, $\theta_e$ is a polar angle between the 
photon initial direction and the normal to the disc and $\pem$ is the azimuthal 
angle, in the disc plane, with respect to the $r$-direction. 

Then, the redshift of the photon emitted from the disc at a distance $r_e$ is given 
by
\begin{equation}
g_{\rm eff}=r_e \left( {\Delta \over A} \right)^{1/2} {(1-V_{\rm K}^2)^{1/2}
\over 1-\omega_{\rm K} \lambda},
\label{g}
\end{equation}
and the angle, $i$, at which the photon will be observed far from the disc in the 
flat space-time is determined by the integral equation of motion
\begin{equation}
\pm \int_{r_e}^\infty \Re^{-1/2}{\rm d} r=\int_{\pi/2}^{i} \Theta^{-1/2} {\rm d} 
\theta.
\label{int}
\end{equation}
where $\Re$ and $\Theta$ are radial and polar effective potentials, respectively,
\begin{eqnarray}
\lefteqn{\Re(r) = (r^2+a^2-\lambda a)^2 - \Delta \left[ \xi^2 + (\lambda - a)^2
\right] ,} \nonumber \\
\lefteqn{\Theta(\theta) =\xi^2 + \cos^2 \theta \left( a^2-\lambda^2/{\sin^2 
\theta} \right).}
\end{eqnarray}

The photon transfer function is tabulated in discrete steps in five dimensions ($a$, 
$r_e$, $\cos i$, $g_{\rm eff}$, $\cos \theta_e$). Generated photons are summed in the 
appropriate elements of the photon transfer function table according to the following 
algorithm.
\begin{enumerate}
\item for given $a$ and $r_e$ generate the photon initial direction from the 
distribution uniform both in $\phi_e$ and $\ct$,  
\item find the constants of motion from equation (\ref{const}), 
\item solve equation (\ref{int}) for $i$ (initial sign of $\Re$ in equation 
(\ref{int}) is negative for $\pi/2 < \phi_e < 3\pi/2$ and positive otherwise),
\item find $g_{\rm eff}$ from equation (\ref{g}),
\item increment an element of the transfer function table corresponding to $a$, 
$r_e$, $\cos i$, $g_{\rm eff}$ and $\cos \theta_e$.
\end{enumerate}
In order to compute and tabulate the transfer function we have calculated 
trajectories of $10^9$ photons. The disc is axially symmetric and emission from a 
given point of the disc is also axially symmetric. Therefore, integration over 
$\phi_e$ for trajectories starting at a given point of the disc and observed at a 
given angle $i$ and collected at all angles $\phi_o$ is equivalent to integration 
over all trajectories starting at a given radius $r_e$, which are observed at a given 
angle $i$ and any particular $\phi_o$.

\section{Transformation of the photon number intensity}
\label{app:intensity_transformation}

As pointed out in Section \ref{sec:model}, application of the photon transfer 
function requires transformation of the photon number intensity between the reference 
frames of the distant observer (the Boyer-Lindquist coordinate frame) and the 
observer co-rotating with the disc. The latter one is represented by the covariant 
tetrad (we show only the components used below):
\begin{eqnarray}
\lefteqn{e^{t'} = - r^{-1}\left( 1-V_{\rm K}^2 \right)^{-1/2} \left({ A \over 
\Delta}\right)^{1/2}} \nonumber \\
\lefteqn{~~~~\times\left( \gtt+\omega_{\rm K} \gtp,0,0,\gtp+\omega_{\rm K} \gpp 
\right),} \nonumber \\
\lefteqn{e^{r'}=(0,r/\sqrt{\Delta},0,0),} \nonumber \\
\lefteqn{e^{\phi'} = r^{-1} \left( 1-V_{\rm K}^2 \right)^{-1/2} \left({ A \over 
\Delta}
\right)^{1/2} \left[ V_{\rm K} \gtt + \gtp \left( {r^2 \Delta^{1/2} \over A} \right. 
\right.} \nonumber \\
\lefteqn{~~~~\left. \left. + {2arV_{\rm K} \over A} \right),0,0,V_{\rm K}\gtp + \gpp 
\left( {r^2 \Delta^{1/2} \over A} + {2arV_{\rm K} \over A} \right)\right],} 
\end{eqnarray}
where order of vector components is given by $(t,r,\theta,\phi)$, and primes denote 
coordinates defined in the disc local rest frame. Then, the one-form ${\rm d} t'$ can 
be written in terms of the Boyer-Lindquist coordinates as
\begin{equation}
{\rm d} t'= e_t^{t'} {\rm d} t + e_\phi^{t'} {\rm d} \phi
\end{equation}
As for the observer at rest in the disc ${\rm d} \phi'=0$, and 
\begin{equation}
{\rm d} \phi'= e_t^{\phi'} {\rm d} t + e_\phi^{\phi'} {\rm d} \phi,
\end{equation}
we obtain the following relation
\begin{equation}
{\rm d} t'=\beta_t {\rm d} t,
\end{equation}
where
\begin{equation}
\beta_t= e_t^{t'}-
e_\phi^{t'} e_t^{\phi'}/e_\phi^{\phi'}.
\end{equation}
Similarly, for the disc unit area we find 
\begin{equation}
{\rm d} r' {\rm d} \phi'=e_r^{r'} {\rm d} r \left( e_t^{\phi'} {\rm d} t+
e_\phi^{\phi'} {\rm d} \phi \right)=\beta_S {\rm d} r {\rm d} \phi
\end{equation}
(in the latter equality we have used the condition ${\rm d} t'=0$,
relevant for the three-space measurements in the disc rest frame), where
\begin{equation}
\beta_S=e_r^{r'}\left( e_\phi^{\phi'} - e_\phi^{t'} e_t^{\phi'}
/e_t^{t'} \right).
\end{equation}
The factor $\beta_S$ gives correction for the integral element ${\rm d} r {\rm d} 
\phi$ in equation (\ref{eq:trans_integ}) (integration over $\phi$ is hidden in the 
convolution) which should be taken into account in calculation of the number of 
photons emitted from the accretion disc.

\label{lastpage}

\end{document}